\documentclass[conference]{IEEEtran}
\IEEEoverridecommandlockouts

\usepackage{cite}
\usepackage{amsmath,amssymb,amsfonts}
\usepackage{graphicx}
\usepackage{textcomp}
\usepackage{booktabs}
\usepackage{siunitx}

\begin{document}

\title{A Reference-Based Protocol for Assessing Image Displacement and Scale Stability}

\author{

\IEEEauthorblockN{1\textsuperscript{st} Fatih Ozturk}
\textit{Yildiz Technical University}\\
\IEEEauthorblockA{\textit{Mechatronics Engineering} \\
Istanbul, Turkey}
\and
\IEEEauthorblockN{2\textsuperscript{nd} Emir Mercanoglu}
\IEEEauthorblockA{\textit{American Collegiate Institute}\\
Izmir, Turkey}
\and
\IEEEauthorblockN{3\textsuperscript{rd} İrem Sayın}
\textit{Yildiz Technical University}\\
\IEEEauthorblockA{\textit{Mechatronics Engineering} \\
Istanbul, Turkey}
\and
\IEEEauthorblockN{4\textsuperscript{th} Huseyin Uvet}
\textit{Yildiz Technical University}\\
\IEEEauthorblockA{\textit{Mechatronics Engineering} \\
Istanbul, Turkey}
}
\maketitle

\begin{abstract}
\normalsize\mdseries
We present a reference-based protocol for assessing image stability,
illustrated with a passive lens support. Pre/post-event wall references
separate image-centre displacement from scale change. All 70 trials
were retained: 30 baseline and 40 supported, acquired sequentially.
Displacement ($p=0.597$) and scale spread ($p=0.356$) showed no
significant difference. Lower median absolute scale deviation in the
supported run persisted with scene-only processing, but its cause
remains unresolved. Marker-estimation eligibility differed completely
between runs for an unknown reason. The contribution is a workflow
that reports complete-run variability, typical deviations and estimator
coverage together; the case study does not establish support efficacy,
equivalence or transient damping.

\end{abstract}
\renewcommand{\IEEEkeywordsname}{Keywords}

\section{Introduction}

Camera pose and imaging geometry affect vision-based displacement
measurements~\cite{b6,b7}. A fixed reference can help distinguish changes
in the camera image from motion of the measured target. Lens and
calibration stability are also relevant~\cite{b2,b3,b4}. Image scale can
change even when displacement at the image centre is small.

We present a reference-based assessment protocol, illustrated with a
printed lens support under impulsive loading. It retains complete trial
sequences, separates displacement from scale, records estimator eligibility
and checks comparisons with a common estimator. These steps expose why
typical deviation, full-run spread and measurement consistency must be
considered together before attributing changes to a mount.

\section{Experimental Setup}

\subsection{Measurement Configuration}

Measurements were performed on a robotic test platform in which a fixed
camera measures target position at a working distance of
3 m. Each trial comprised mechanically drawing and releasing a bow
to launch an arrow, using the same draw-and-release procedure and
settings in both conditions. One frame was acquired before release and
one afterwards. Acquisition intervals were not measured and were only
approximately consistent across trials; the frame pairs do not resolve
the intervening transient.

Focus, zoom, exposure and gain were locked for
all trials; automatic focus and automatic gain were disabled.

\subsection{Fiducial Reference}

Two ArUco markers~\cite{b1} (\texttt{DICT\_7X7\_50}, IDs 0 and 1) of 110 mm side
length were placed in the camera field of view, one in the upper left
and one in the lower right of the frame. The markers were fixed directly
to the laboratory wall, mechanically decoupled from the target and its
stand, to provide a stationary reference independent of target motion. Marker distance from the camera was 4.0 m, measured
directly.

The effective focal length was recovered from the imaged marker size,
\begin{equation}
f = \frac{L_{\text{px}} \, Z}{L_{\text{mm}}},
\label{eq:focal}
\end{equation}
giving $f = 7102$ px and an angular scale of 0.141 mrad
per pixel.

The Raspberry Pi HQ Camera (Sony IMX477, rolling shutter) was operated
at $2028\times1520$ resolution and fitted with a Canon EF-S 18--135 mm
zoom lens via an adapter. Combined camera--lens mass was 530 g.

\begin{figure*}[t]
\centering
\includegraphics[width=0.72\textwidth]{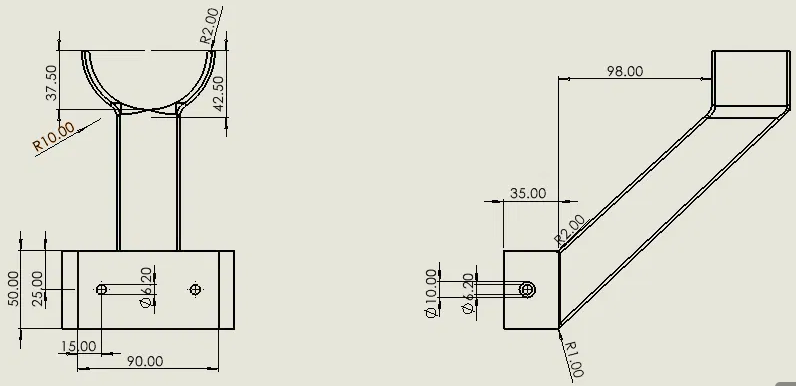}\hfill
\begin{minipage}[b]{0.24\textwidth}
\centering
\includegraphics[width=0.65\linewidth]{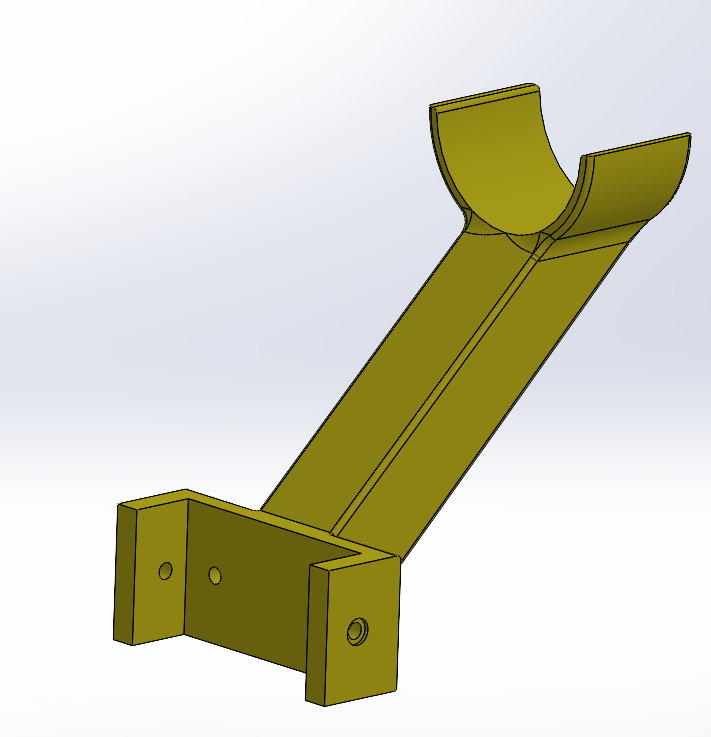}\par\vspace{3pt}
\includegraphics[width=\linewidth]{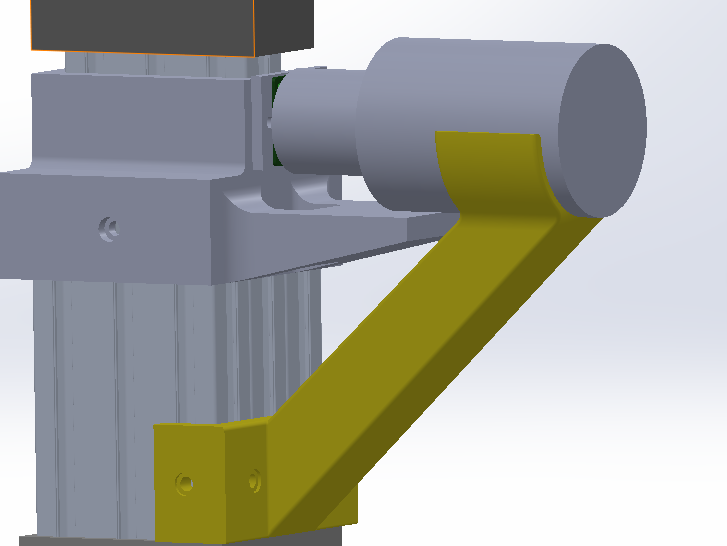}
\end{minipage}
\caption{Proposed lens-support bracket: dimensioned front and side views
(left; dimensions in millimetres), isolated 3D view (upper right), and
installed CAD configuration (lower right). The highlighted bracket
supports the underside of the lens.}
\label{fig:cad}
\end{figure*}

\section{Mount Design}

The proposed bracket was designed to limit lens motion during impulsive
loading without changing the camera-body fixing. The single-piece
design combines a semicircular cradle that grips the lens barrel from
below with an inclined support arm, as shown in Fig.~\ref{fig:cad}.
The arm terminates in a mounting plate shaped to engage the side
openings of the existing bracket. The lens fits into the cradle without
retaining bolts, allowing the camera to be removed while the support
remains installed.

The bracket was printed in PLA with 25\% gyroid infill and two perimeter
walls. Infill geometry can affect printed-polymer damping~\cite{b8,b9},
but damping properties of this bracket were not measured.

\section{Method}

\subsection{Trial Protocol}

All 30 baseline and 40 supported frame pairs were analyzed, including
the first five in each run; none were removed as settling trials or
outliers. Ten no-event pairs characterized static repeatability.

Both runs were acquired on the same day with unchanged camera and
loading settings. The presence of the lens support was the only
planned mechanical change; this does not establish identical image
quality or estimator eligibility between runs. Session records place the supported
run first (14:21--14:47), followed by the baseline run (15:21--15:46).
The conditions were not interleaved.

\subsection{Transform Estimation}
\label{sec:estimator}

Each image pair was fitted with a similarity transform and decomposed
into scale $s$, rotation $\theta$ and translation. With two marker IDs detected in both frames, the eight sub-pixel corners were fitted using least median of
squares. Otherwise, SIFT~\cite{lowe2004} matches from the static background were fitted
using RANSAC (1.5 px reprojection threshold), with the target and stand
masked out. All 30 baseline pairs used scene features; all 40 supported
pairs and 10 static pairs used markers. Baseline logs record zero common
marker IDs in 25 pairs and one in five; none met the two-common-marker
requirement. These counts concern shared IDs, not detection in individual frames.
The cause of failure is unknown; visibility, image quality and detector
behaviour were not independently resolved.

Reprocessing all 40 supported pairs with scene features gave mean
absolute differences of 0.09 px in displacement and 0.0001 in scale.
The complete-run comparisons remained non-significant for displacement
($p = 0.597$) and scale spread ($p = 0.354$), with an SD ratio of
0.931. These two test conclusions were unchanged, but this check does not
exclude image-dependent measurement bias or run-order confounding.

\subsection{Reported Quantities}

Two quantities are reported per trial. \emph{Displacement} is the motion
of the image centre under the estimated transform. Its angular
equivalent is $d/f$; multiplying by distance gives an apparent
position error under the small-angle approximation.
\emph{Scale deviation} is $s - 1$; because a scale change displaces
features in proportion to their radius from the image centre, its
effect is reported at the corner radius $r=1267$ px. The principal
point is approximated by the image centre. At distance $Z$, its apparent
position-error SD is $Zr\,\mathrm{SD}(s-1)/f$. Displacement is summarized by median
and interquartile range (Q1--Q3), with mean $\pm$ standard deviation
also reported. Scale variability is summarized by the standard deviation
of $s-1$; median and IQR of $|s-1|$ additionally describe typical
scale-deviation magnitudes. All comparisons use the full runs.
The two-sided Mann--Whitney test uses a continuity-corrected asymptotic
$p$ value. Displacement effects are defined as proposed minus baseline;
the Hodges--Lehmann estimate is the median of all between-group pairwise
differences, with a 95\% confidence interval obtained by exact rank-sum
inversion under a common-shape location-shift model. The scale SD ratio
is baseline divided by proposed; its percentile bootstrap interval uses
10,000 independent within-condition resamples (seed 20260919).
Levene's test is centred on the group medians.
An additional post-hoc analysis compares $|s-1|$ distributions using
Mann--Whitney and estimates the difference of sample medians with
10,000 independent within-run percentile bootstrap resamples
(seed 20260922). It is repeated with scene features for both runs;
these exploratory $p$ values are unadjusted and are not tests solely
of median equality.

\subsection{Static Repeatability}

Under static conditions the method returned a displacement of
0.17 $\pm$ 0.06 px (maximum
0.24 px) and a scale standard deviation of
0.00003. These values characterize marker-based repeatability, not a
calibrated detection limit.

\section{Results}

\subsection{Trial Sequence}
\label{sec:sequence}

\begin{figure}[t]
\centerline{\includegraphics[width=\columnwidth]{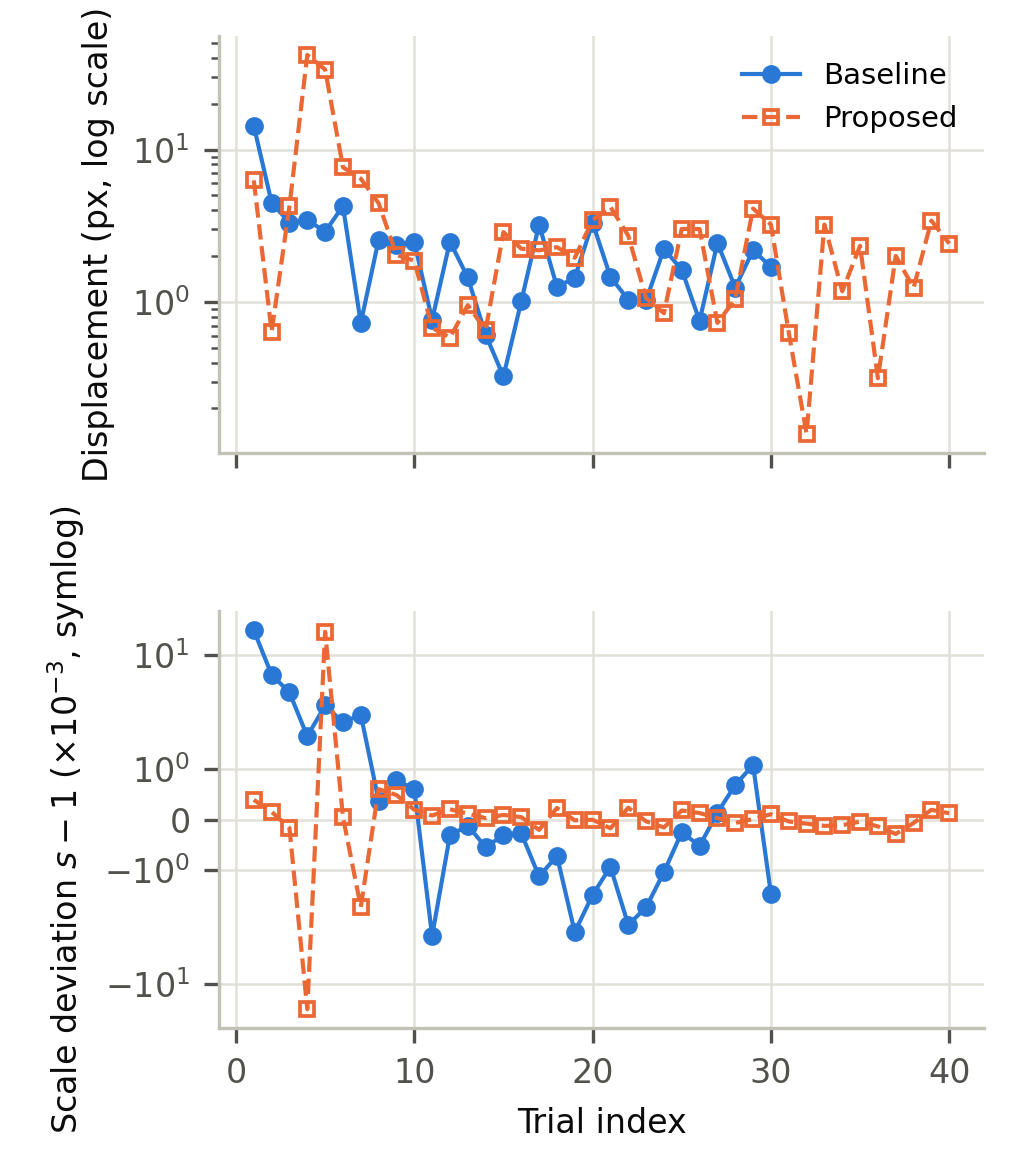}}
\caption{Displacement and scale deviation against trial index for both
conditions. All 30 baseline and 40 supported trials are included.
Displacement uses a logarithmic axis; signed scale deviation uses a
symmetric logarithmic axis, linear within $\pm2\times10^{-3}$.}
\label{fig:sequence}
\end{figure}

Initial deviations were larger (Fig.~\ref{fig:sequence}): first-five mean
displacements were 17.31 px with support and 5.70 px without (maxima
41.82/14.35 px). Settling is a possible but unverified explanation.
These trials remain in every summary and test.

\subsection{Displacement}

\begin{figure}[t]
\centerline{\includegraphics[width=\columnwidth]{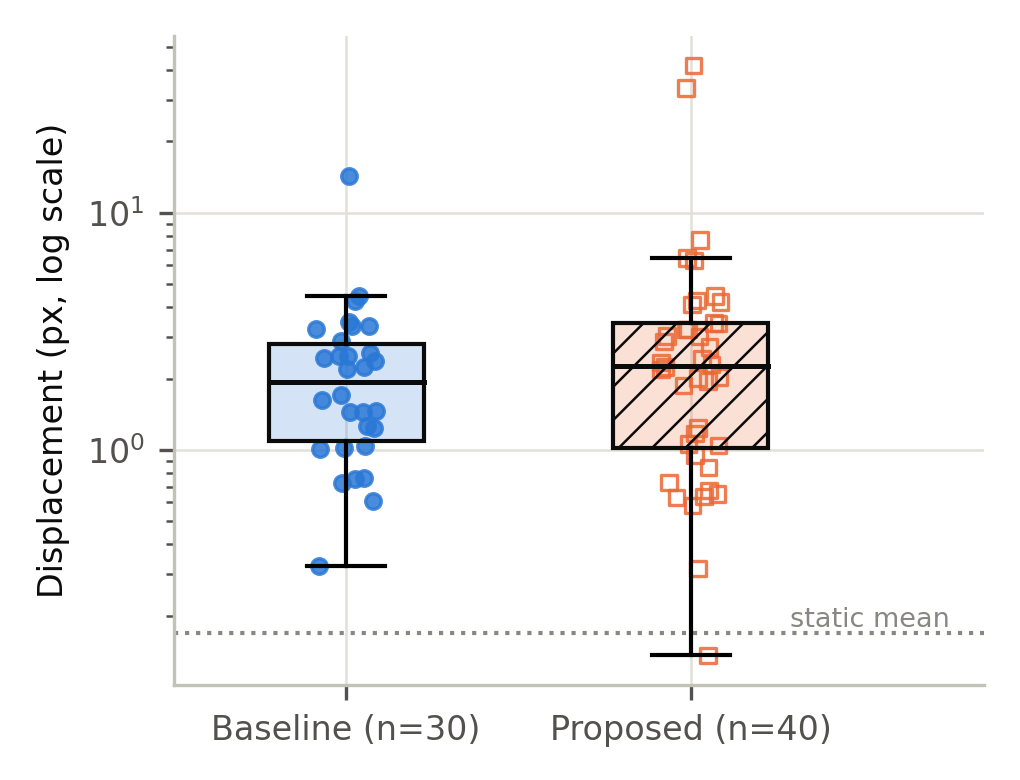}}
\caption{Complete-run displacement distributions on a logarithmic axis.
All observations are shown; the horizontal line marks mean static
displacement. Boxes show Q1--Q3 with the median; whiskers extend to
the most extreme observations within 1.5 IQR of the quartiles.}
\label{fig:displacement}
\end{figure}

No statistically significant image-centre displacement difference was
detected between the two conditions. Median displacement was 2.26 px
(IQR 1.02--3.42) with support and 1.94 px (IQR 1.09--2.80) without
support (means $4.19\pm8.03$ and $2.41\pm2.51$ px, respectively).
A Mann--Whitney $U$ test gave $p = 0.597$, with a rank-biserial
correlation of 0.075 and a Hodges--Lehmann shift estimate of 0.226 px
(95\% CI $-0.399$ to 1.034 px; support minus baseline).
Fig.~\ref{fig:displacement} includes the initial deviations;
the interval spans both directions and does not establish equivalence.

\subsection{Scale Stability}

\begin{figure}[t]
\centerline{\includegraphics[width=\columnwidth]{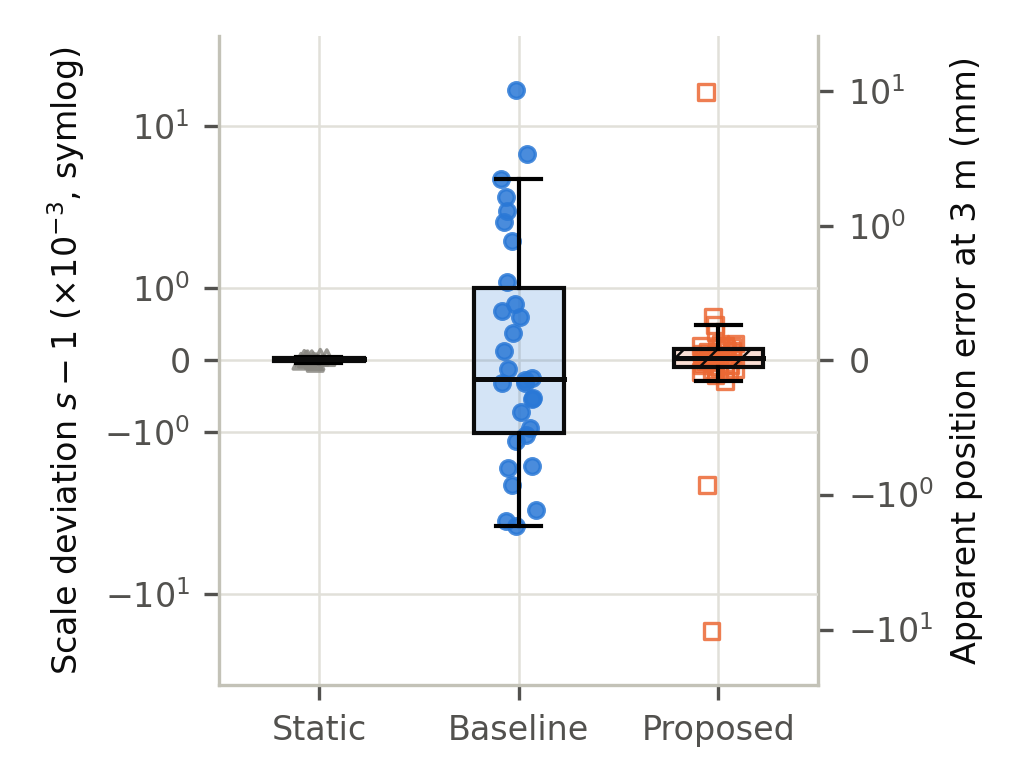}}
\caption{Complete-run scale deviations and static reference measurements.
The left axis is symmetric logarithmic, linear within
$\pm2\times10^{-3}$; all observations are shown. Box and whisker
definitions match Fig.~\ref{fig:displacement}. The right axis gives
the equivalent apparent position error at 3 m.}
\label{fig:scale}
\end{figure}

The standard deviation of $s-1$ was 0.00394 without support and
0.00424 with support (Fig.~\ref{fig:scale}). The baseline/support SD
ratio was 0.929 (bootstrap 95\% CI 0.275--19.370).
Levene's test gave $p = 0.356$; it tests spread rather than central
tendency. No statistically significant spread difference was detected,
and the wide interval leaves the SD ratio poorly determined.
Median $|s-1|$ was $1.063\times10^{-3}$ at baseline and
$0.127\times10^{-3}$ with support (Table~\ref{tab:summary}). The post-hoc
Mann--Whitney comparison gave $p=3.30\times10^{-8}$. The median
difference (support minus baseline) was $-0.936\times10^{-3}$,
with a percentile bootstrap 95\% CI of
$[-1.573,-0.489]\times10^{-3}$. With scene features in both runs,
the supported median was $0.100\times10^{-3}$ and
$p=4.05\times10^{-8}$. Thus, the lower typical deviation survives
this estimator check, while large initial deviations still influence
the full-run SD. It could reflect a physical difference or an acquisition or measurement artefact; neither analysis identifies a causal support effect.
At $r=1267$ px, the scale SDs correspond to apparent position-error
SDs of 2.11 mm without support and 2.27 mm with support at 3 m.
These geometric equivalents are not measured target errors.

\begin{table}[t]
\caption{Complete-run measurements and static reference; no trials excluded.}
\label{tab:summary}
\centering
\small
\setlength{\tabcolsep}{3pt}
\begin{tabular}{lccc}
\toprule
 & Static & Baseline & Proposed \\
\midrule
$n$ & 10 & 30 & 40 \\
Displacement, median (px) & 0.20 & 1.94 & 2.26 \\
\quad Q1--Q3 (px) & 0.12--0.21 & 1.09--2.80 & 1.02--3.42 \\
Displacement, mean (px) & 0.17 & 2.41 & 4.19 \\
\quad SD (px) & 0.06 & 2.51 & 8.03 \\
\quad as mm @ 3 m & 0.07 & 1.02 & 1.77 \\
Scale SD ($\times10^{-3}$) & 0.03 & 3.94 & 4.24 \\
\quad as mm @ 3 m & 0.01 & 2.11 & 2.27 \\
$|s-1|$, median ($\times10^{-3}$) & 0.02 & 1.06 & 0.13 \\
\quad Q1--Q3 ($\times10^{-3}$) & 0.02--0.03 & 0.53--2.15 & 0.06--0.20 \\
\bottomrule
\end{tabular}
\end{table}

\subsection{Illustrative Sensitivity}

For independent normal samples with $n=30/40$, common SD 8 px,
and two-sided $\alpha=0.05$, noncentral-$t$ calculation gives an
80\%-power mean difference of 5.49 px (standardized effect 0.686).
Using the observed SDs of 2.51/8.03 px gives an approximate Welch
value of 3.86 px. These illustrative mean-difference models do not
estimate the power of the rank test on skewed, possibly dependent
trials or bound the true effect. No equivalence margin was specified.

\section{Discussion}

The protocol distinguishes lower typical absolute scale deviation
from unchanged full-run spread and exposes estimator coverage as a
diagnostic. The exploratory absolute-scale association survived the
common-estimator check; nevertheless, it could reflect a physical
change or a measurement artefact. This worked procedure is not an independently validated accuracy
standard or evidence of bracket efficacy.

Axial camera translation and changes in lens geometry can both produce
scale variation at the wall reference plane; these data cannot separate
them. Excitation transmission was not measured. Image-centre motion
also omits position-dependent rotation and scale components, so scale
SD cannot be added to mean displacement to infer total target error.

\subsection{Limitations and Further Work}
\label{sec:limitations}
Condition and run order are confounded despite same-day acquisition
and unchanged settings. Tests and bootstrap intervals assume independent,
exchangeable trials; serial dependence and changing variability can
invalidate their uncertainty estimates. All inferences are exploratory.
Unknown baseline marker-detection failure leaves possible systematic
acquisition or measurement differences unresolved, even with common
scene-feature processing. Static repeatability covers markers only;
calibration uncertainty was not propagated into millimetre equivalents.

Unmeasured frame intervals prevent checking a common stage of transient
decay. No damping ratio, settling time or rolling-shutter immunity is
established. Repeated installations with randomized condition order,
consistent estimator coverage and controlled timing are needed for
causal comparisons. References at multiple depths, independent motion
sensing and time-resolved acquisition would help distinguish optical
changes, camera motion and transient response.

\section{Conclusion}

This 70-trial case study illustrates a reference-based assessment
workflow. Displacement ($p=0.597$) and scale spread ($p=0.356$)
showed no significant difference. Lower typical absolute scale deviation
persisted with scene-only processing, but sequential runs and unexplained
marker-detection failure prevent attributing it to the support.
Neither support efficacy nor equivalence is established.


\end{document}